# Reentrant phase in nanoferroics induced by the flexoelectric and Vegard effects

Anna N. Morozovska[1] and Maya D. Glinchuk [2*],

[1] Institute of Physics, National Academy of Sciences of Ukraine,

46, pr. Nauky, 03028 Kyiv, Ukraine

[2] Institute of Problems for Material Sciences, National Academy of Sciences of Ukraine,

3, Krjijanovskogo, 03068 Kyiv, Ukraine

**Abstract**

We explore the impact of the flexoelectric effect and Vegard effect (chemical pressure) on the phase diagrams, long-range polar order and related physical properties of the spherical ferroelectric nanoparticles using Landau-Ginzburg-Devonshire phenomenological approach. The synergy of these effects can lead to the remarkable changes of the nanoparticles' phase diagrams. In particular, a commonly expected transition from ferroelectric to paraelectric phase at some small critical size is absent; so that the critical size loses its sense. Contrary, the stabilization of the ferroelectric phase manifests itself by the enhancement of the transition temperature and polarization with the particle size decrease. Ferroelectric phase reentrant phenomenon was observed earlier in the tetragonal $BaTiO_3$ nanospheres of radii $5 - 50$ nm [Zhu et al., JAP **112**, 064110 (2012)] and stayed unexplained up to now. Our calculations have shown the physical mechanism of the exciting phenomenon is the flexo-chemo-effect. Since the spontaneous flexoelectric coupling, as well as ion vacancies, should exist in any nanostructured ferroelectrics, obtained analytical results can be valid for many nanoferroelectrics, where reentrant phases appearance can be forecasted.

---

[*] Corresponding author: glim1@voliacable.com

# I. Introduction

Unique physical properties of nanosized multiferroics attract the permanent attention of researchers [1]. Many experimental and theoretical studies of nanograined ceramics or nanopowders conclude that the transition from the ferroelectric, (ferromagnetic) phase to the paraelectric (paramagnetic) phase occurs at some critical size with the size decrease. Using the conception of the critical size within the theory of size effects, an excellent ability to manage the temperature of the ferroelectric phase transition, the magnitude and position of the maximum of the dielectric susceptibilty and other properties for many different forms of nanoparticles [1]. In particular, it has been demonstrated for spherical nanoparticles and nanograined ceramics of $BaTiO_3$ and $PbTiO_3$ [2, 3, 4]. Disappearance of ferroelectric phase for the sizes smaller than critical one seems to be in agreement with fundamental physics statement about necessity to have large amount of interacting ions for creation of mean field as the characteristic feature of a long-range order (so called correlation effect) [5]. Theoretical consideration of size effects allows one to estimate the critical size, establish the physical origin of transition temperature shift relatively to the bulk value, as well to calculate the changes of phase diagrams appeared under decreasing of nanoparticles sizes and shape. In particular, using the continual phenomenological approach Niepce [6], Huang et al [7, 8], Ma [9] and Morozovska et al [10, 11, 12, 13, 14, 15] have shown, that the critical sizes existence, size-driven changes of the transition temperatures leading to the enhancement or weakening of polar properties are conditioned by the different shape, surface and strain effects in nanoparticles. In particular, there are **several intriguing cases.**

The conservation of polar properties was discovered by Yadlovker and Berger [16, 17, 18] in cylindrical nanoparticles of Rochelle salt with radius less than 15 nm at temperatures higher than the bulk Curie temperature up to the decomposition temperature. The behaviour .was theoretically explained in Ref.[10, 12, 14] on the base of the ordering role of the biaxial ***intrinsic surface stress*** (surface tension) [19, 20] in nanocylinders, while the hydrostatic pressure induced by the surface tension in nanospheres [13, 15] as well as the bond contraction [7, 8] can only increase the critical size. The correlation and depolarization effects can only increase the critical size [10-15] and so they decrease the ferroelectric phase region.

Zhu et al [21] observed the appearance of reentrant tetragonal phase in $BaTiO_3$ nanospheres at room temperature along with the enhancement of polarization and transition temperature for the particle sizes less than 20 nm and narrow distribution of the particle sizes. Therefore the transition to cubic phase is absent and the critical size does not exist for the case. As it was shown earlier [22], the size-induced phase transition from a ferroelectric to paraelectric phase disappears in ferroelectric nanopills and nanowires due to the spontaneous flexoelectric effect. However, the physical nature of the reentrant phase stays unclear up to now, although its understanding can be very important both for fundamental science and for practical applications. This problem solution will be useful not only for BaTiO$_3$ nanoparticles, but practically for other nano-ferroelectrics. Therefore the elaboration of the theoretical explanation that can pour light on physical mechanisms of reentrant phase appearance in nano-ferroelectric materials is the main goal of the work.

We explore the impact of the *flexoelectric effect* and *Vegard effect* (*chemical pressure*), intrinsic surface stresses, correlation and depolarization effects on the phase diagrams, long-range polar order and related physical properties of the spherical ferroelectric nanoparticles using Landau-Ginzburg-Devonshire phenomenological approach. Below we will show both effects can be a decisive physical mechanism ruling the observed phenomena.

*Flexoelectric effect*, firstly predicted theoretically by Mashkevich and Tolpygo [23] in 1957, exists in any material, making the effect universal [24, 25, 26, 27]. The flexoelectricity impact is of great importance in nanoscale objects, for which the strong strain gradients are inevitable present near the surfaces, in thin films [28, 29, 30], at the domain wall or ferroelectric interfaces [29, 31, 32, 33], around point and topological defects [25, 26]. Owing to the surface or interface effects, flexoeffect appears spontaneously near the surface of any nanoparticle [22], everywhere where the spontaneous polarization distribution becomes inhomogeneous. Notably, that the influence of flexoelectricity is important not only in thin films and nanoparticles, but also in micro- and nanograined ceramics [34, 35]. Despite the great importance of the flexoelectricity, its tensorial components strength and manifestations, as estimated by Kogan [36] in early 1963, remained poorly known for most of the ferroic materials, except for the experimental measurements of some components for ferroelectric perovskites [37, 38, 39] and *ab initio* calculations [40, 41]. However all of these experimental and theoretical results do not allow obtaining the complete information about the full tensor and are rather contradictory [42], indicating on a limited understanding of the effect nature. Overall, much more is predicted theoretically than found experimentally.

The nature of the *chemical pressure effect* [43], that other names are compositionally induced *Vegard strains* [44, 45] or elastic dipole [45], is the following. Any point defect (interstitial atoms, impurities and vacancies) leads to the local deformation of the crystal lattice, and the action of many defects causes the strain proportional to their concentration [45, 46, 47]. The influence of Vegard strain, coming from the diffusion and the accumulation of defects near the interfaces, results in the pronounced change of their polar properties [48, 49]. Unfortunately exact values of the Vegard strain tensor components are purely known, but its order of magnitude is available for perovskites [45].

## II. Analytical theory

LGD-functional bulk (b) and surface (s) energy densities for a nanoparticle with uniaxial ferroelectric polarization $\vec{P} = (0,0,P_3)$ acquires a relatively simple form [22]:

$$\Phi_b = \int_V d^3r \left( \begin{array}{c} \alpha_b(T)\dfrac{P_3^2}{2} + \beta\dfrac{P_3^4}{4} + \gamma\dfrac{P_3^6}{6} + \dfrac{g_{ij33}}{2}\left(\dfrac{\partial P_3}{\partial x_i}\dfrac{\partial P_3}{\partial x_j}\right) - \dfrac{F_{3ijk}}{2}\left(P_3\dfrac{\partial \sigma_{ij}}{\partial x_k} - \sigma_{ij}\dfrac{\partial P_3}{\partial x_k}\right) \\ -P_3\left(\dfrac{E_3^d}{2} + E\right) - Q_{ij33}\sigma_{ij}P_3^2 - \dfrac{s_{ijkl}}{2}\sigma_{ij}\sigma_{kl} - W_{ij}\sigma_{ij}\delta N \end{array} \right) \quad (1a)$$

$$\Phi_S = \int_S d^2r \left( \dfrac{\alpha^S}{2}P_3^2 + \dfrac{\beta^S}{4}P_3^4 + \mu^S_{\alpha\beta}u_{\alpha\beta} + \ldots \right) \quad (1b)$$

The coefficient $\alpha_b(T)$ typically depends on the temperature $T$. Here we assume the linear dependence, $\alpha_b(T) = \alpha_T(T - T_C)$, where $T_C$ is a Curie temperature. Coefficients $\beta$ and $\gamma$ are regarded temperature independent, $\gamma > 0$. $P_3$ stands for the ferroelectric polarization, $E_3^d$ is depolarization field, $Q_{ijkl}$ are the bulk electrostriction tensor coefficients, $g_{ijkl}$ is the gradient coefficients tensor, $F_{ijkl}$ is the flexoelectric stress tensor, $\sigma_{ij}$ is the stress tensor, $W_{ij}$ is the elastic dipole (or Vegard strain) tensor, $\delta N = N(\vec{r}) - N_e$ is the difference between the concentration of defects $N(r)$ in the point **r** and their equilibrium (average) concentration $N_e$. Hereinafter the Vegard tensor is regarded diagonal, i.e. $W_{ij} = W\delta_{ij}$ ($\delta_{ij}$ is delta Kroneker symbol).

Coefficients $\alpha^S$ and $\beta^S$ are supposed to be positive and temperature independent, $\mu^S_{\alpha\beta}$ is the surface stress tensor [20, 23], $u_{ij}$ is the stress tensor.

As one can see from Eq.(1a), the coefficient $\alpha_b(T)$ becomes renormalized by elastic stresses via electrostriction. More rigorously, direct variational method shows that renormalized coefficient $\alpha$ depends on temperature, particle shape and sizes, polarization orientation, correlation and depolarization effects [10-15]. Below we will calculate the renormalized coefficient using definite models for elastic stresses and depolarization field acting on the inner part of a spherical particle of radius $R$ covered with an ultra-thin defect layer of thickness $R_0$ ($R_0 \ll R$). Most of defects are located in the ultra-thin layer and their concentration decreases exponentially towards the particle bulk [51]. The "core" of radius $R - R_0 \approx R$ is free from defects, i.e. $\delta N = 0$ in the core and $\delta N = N$ in the shell. [**Figure 1**]

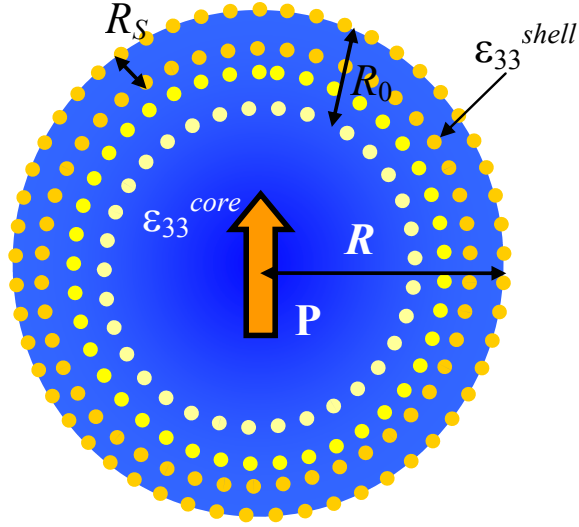

**Figure 1.** Schematics of the spherical particle with radius $R$ covered by the shell of thickness $R_0$ with accumulated defects. Separation between the screening charge and the spontaneous polarization abrupt at the core-shell interface is $R_S$.

In particular we assume that the flexoelectric effect, intrinsic surface stress, bond contraction and Vegard chemical pressure contribute into the elastic stress and strain field under the curved surface of the particle. Acting elastic fields distribution strongly depends on the mechanical boundary conditions, which corresponds to the fixed normal stresses at the free surfaces and displacements at the clamped surfaces of the particle. Additional terms in the boundary conditions can originate from the flexocoupling energy [22, 50].

**A. Flexoelectric stresses.** General expressions for the flexoelectic stresses have the form:

$$\delta\sigma_{ij} = F_{ijkl}\frac{\partial P_k}{\partial x_l}. \qquad (2)$$

The strain induced by the spontaneous polarization gradient we are interested in, are noticable in near the surface, where the gradient is essential. Since we consider a sphere within core-and-shell model [6, 10] and regard that the shell thickness $R_0$ is much thinner than the core radius $R - R_0$, we can approximate the flexoelectric strain value in every small part of the shell by the ones of the thin film or pill derived by Eliseev et al [22]. Thus we can use the expressions [22] for the flexoelectric strains, $\delta u_{rr} = -(f_{11}/c_{11})\partial P_3/\partial r$ and $\delta u_{r\theta} = -(f_{44}/2c_{44})\partial P_3/\partial r$ ($P_3 = P_S \cos\theta$), as estimation in the local reference frame. Here $c_{ij}$ are elastic stiffness, $f_{ij}$ are the flexoelectric strain coefficients.

**B. Vegard-type chemical stresses.** Defects accumulation under a curved surface produces effective stresses of the inner part of the particle due to the lattice expansion or contraction [51,15]. The characteristic thickness of the layer enriched by defects is determined by the screening length, and their maximal concentration is limited by steric effect [48, 49]. Hence the hydrostatic pressure excess in the inner part of the particle can originate from the surface tension, bond contraction and Vegard mechanisms. For a spherical particle of radius $R$ the pressure excess acquires the form [13, 15]:

$$\delta\sigma_{rr}(R) = -\frac{2\mu}{R} - \frac{\eta}{s_{11} + 2s_{12}}\frac{R_0^2}{R^2}. \tag{3}$$

Here subscript "$rr$" denotes that the pressure is radial. $R$ is the particle radius, $\mu$ is the surface tension coefficient, $s_{ij}$ are elastic compliances modulus of the material, the characteristic size $R_0$ is the particle surface layer thickness, where accumulated defects create elementary volume changes. Parameter $\eta \equiv W\delta N$ is a "compositional" Vegard strain. For perovskites $ABO_3$ the absolute values of $W$ related with vacancies can be estimated as $|W| \propto (10 - 10^2)$ Å$^3$ [45] and so corresponding Vegard strains $\delta u_{ij} = W_{ij}\delta N$ can reach percents for relatively high variation of defect concentration $\delta N \sim 10^{27}$ m$^{-3}$.

**C. Depolarization field model.** Let us estimate the depolarization field existing inside a singe-domain spherical nanoparticle core. The core ferroelectric polarization is $\vec{P} = (0,0,P_3)$, relative dielectric permittivity is $\varepsilon_{33}^{core}$ and free carriers are regarded absent. The core is covered by a non-ferroelectric paraelectric shell with relative dielectric permittivity $\varepsilon_{33}^{shell}$, at that the strong inequality $\varepsilon_{33}^{shell} \gg \varepsilon_{33}^{core}$ is likely when the shell is regarded to be in a paraelectric phase close to the ferroelectric transition. The screening charge is located either immediately outside the shell or near its outer surface, so the "effective" separation between the screening charge and the spontaneous polarization abrupt at the core-shell interface is $R_S$ and $R_S \leq R_0$ (see **Figure 1**). The electrostatic problem can be solved exactly for a spherical particle:

$$E_3^d = \frac{-\left(R^3 - (R-R_S)^3\right)\left(P_3 - \overline{P_3^{shell}}\right)}{\varepsilon_0\left((R-R_S)^3\left(\varepsilon_{33}^{shell} - \varepsilon_{33}^{core}\right) + R^3\left(2\varepsilon_{33}^{shell} + \varepsilon_{33}^{core}\right)\right)}. \tag{4}$$

Under the typical condition $R_S \ll R$ the expression can be approximated as $E_3^d \approx -\frac{2R_S\left(P_3 - \overline{P_3^{shell}}\right)}{3\varepsilon_0\varepsilon_{33}^{shell}R}$. If the shell is semiconducting the value $R_S$ acquires the sense of the Tomas-Fermi screening radius and its value can be much smaller than a lattice constant [52].

**D. Size effect of the transition temperature, spontaneous polarization and tetragonality.** Using expressions for elastic stresses (2)-(3) and depolarization field (4) in a spherical particle, the renormalization of the coefficient α for the first order phase transitions, Curie ($T_C$) and FE transition ($T_{FE}$) temperatures acquire the form:

$$\alpha = \alpha_T \left(T - T_C^b\right) + \left(\frac{A}{R} + \frac{B}{R^2}\right) \quad (5)$$

$$T_C(R) = T_C^b - \frac{1}{\alpha_T}\left(\frac{A}{R} + \frac{B}{R^2}\right), \quad T_{FE}(R) = T_\theta^b - \frac{1}{\alpha_T}\left(\frac{A}{R} + \frac{B}{R^2}\right) \quad (6)$$

Where $T_C^b + \frac{3\beta^2}{16\gamma\alpha_T} \leq T_\theta^b \leq T_C^b + \frac{\beta^2}{4\gamma\alpha_T}$ due to the temperature hysteresis [53]. The parameters:

$$A = \frac{2R_S}{3\varepsilon_{33}^{shell}\varepsilon_0} + \frac{2g_{11}^* + 4g_{12}^*}{(R_z + \lambda^*)} + (4Q_{12} + 2Q_{11})\mu, \quad (7a)$$

$$B = \frac{(4Q_{12} + 2Q_{11})\eta R_0^2}{(s_{11} + 2s_{12})} - \frac{12f^2 R_z}{(R_z + \lambda^*)} \quad (7b)$$

The renormalized gradient coefficients $g_{11}^* = g_{11} - f_{11}^2/c_{11}$, $g_{12}^* = g_{12} - f_{44}^2/c_{44}$, extrapolation length $\lambda^* = \frac{g_{11}}{\alpha^S}\left(1 - \frac{f_{11}^2}{2c_{11}g_{11}}\right)$ and correlation length $R_z = \sqrt{g_{11}^*\varepsilon_0\varepsilon_b}$ are introduced in Eqs.(7) [22, 54]. Flexo-parameter $f$ is proportional to the components of the flexoelectric tensor $f_{ij}$ and elastic stiffness $c_{ij}$. Expression for $f$ is listed the **Appendix A** in the **Suppl.Mat.** [55].

As one can see from Eqs.(5)-(7) the size-dependent shift of Curie and ferroelectric phase transition temperatures is determined by the contributions of depolarization field (the first term in the parameter $A$); correlation effect (the second term in $A$); surface tension (the third term in $A$); Vegard strain (the first term in the parameter $B$) and the flexoelectric effect (the second term in the parameter $B$). Equations (5)-(7) show that Curie and transition temperatures $R$-dependences are governed by the signs and values of parameters $A$ and $B$. That constant $A$ is always positive for all known ferroics with perovskite structure, because $2Q_{12} + Q_{11} > 0$ for the materials and μ should be positive for the surface stability, depolarization and correlation terms are also positive. The parameter B can be positive, zero or negative, because the sign of the Vegard strain η is not fixed. In particular it is positive for high enough tensile Vegard strains η>0 and relatively small flexo-parameter $f^2$, because its contribution to Eq.(7b) is always negative. The condition $B<0$ is deliberately true for zero or compressive strains η≤0.

The type of the phase diagram depends essentially on the critical radius (none, the single one $R_{cr}$, or two ones $R_{cr}^{min}$ and $R_{cr}^{max}$) that can be obtained from the solution of equation $T_{FE}(R_{cr}) = 0$ (see details in the **Appendix B** in [55]). For the case $A>0$ and $B>0$ the only critical radius exists, $R_{cr} = (D+A)/2T_\theta^b \alpha_T$, parameter $D = \sqrt{A^2 + 4BT_\theta^b \alpha_T}$. The size-induced transition to the paraelectric (PE) phase occurs at $R<R_{cr}$. Under the conditions $A>0$, $B<0$ and $A^2 + 4BT_\theta^b \alpha_T > 0$, there exist two critical radii, $R_{cr}^{min} = (A-D)/2T_\theta^b \alpha_T$ and $R_{cr}^{max} = (D+A)/2T_\theta^b \alpha_T$. Correspondingly FE to PE phase transition occurs at $R_{cr}^{max}$ and PE to FE transition at $R_{cr}^{min}$. Under the conditions $B<0$ and $A^2 + 4BT_\theta^b \alpha_T < 0$ the critical radius does not exist and the nanoparticle maintains its ferroelectric state up to the ultra-small sizes. Exactly the reentrant ferroelectric phase appear in the case and enhances at small $R$. Since both conditions $B<0$ and $A^2 + 4BT_\theta^b \alpha_T < 0$ can be valid for relatively high flexo-parameter $f$ and/or compressive Vegard strains $\eta<0$, only these two mechanisms can enhance the ferroelectric properties of nanospheres (in contrast to the nanowires, where the surface tension can maintain and improve the properties due to the condition $Q_{12} < 0$ [10]).

The spontaneous polarization is $P_S^2(R,T) = \left(\sqrt{\beta(T)^2 - 4\alpha(R,T)\gamma(T)} - \beta(T)\right)/2\gamma(T)$. Ratio of the lattice constants $c/a$ reflecting the tetragonality is $1 + (Q_{11} - Q_{12})k^2 P_S^2$, where the coefficient $k = \langle\cos\theta\rangle$ reflects the random disordering of the crystallographic axes in the ensemble of nanoparticles or ceramic grains orientation in the ordered FE phase, For the case of tetragonal or rhombohedral symmetry of the grains $k=0.831$ or $k=0.866$ correspondingly [56, 57].

### III. Reentrant phase in the spherical BaTiO$_3$ nanoparticles: theory and experiment

Since the parameters μ, η and $f$ are included in the expressions for $A$ and $B$ in an additive way, only their combinations can be reliably defined from the fitting to experimental data. Below we demonstrate the possibility for the spherical BaTiO$_3$ particles. The known parameters α, $\beta_{ij}$, $\gamma_{ijk}$, $Q_{ij}$, $s_{ij}$, $c_{ij}$, $g_{ij}$, $f_{ijkl}$, $F_{ijkl}$ and $R_z$ of the bulk BaTiO$_3$ along with the reasonable ranges of the surface influence related parameters $R_0$, $R_S$, $f$, λ, μ and η are listed in the **Table SI** in **Appendix C** [55]. Note that we chose very small "seeding" values of λ in order to show clearly the influence of the flexocoupling mechanism. The chosen value $R_S = 0.1$ nm corresponds to a rather small, but reasonable depolarization effect. $R_0$ varies in the range (0.8 – 4) nm depending on the defect concentration gradient.

Dependence of the transition temperature $T_{FE}$ on the particle radius was calculated at different flexo-parameter $f$ and Vegard strain η. Results are shown in the **Figures 2a-2b.**

Different curves 1-4 in the **Figure 2a** correspond to zero Vegard strain (η=0) and different effective flexo-parameter $f$, $f_1<f_2<f_3<f_4$. The ferroelectric transition temperature monotonically increases with particle radius increase for the smallest value $f_1=2.01\times10^{-6}$ V/Pa (curve 1 in the **Figure 2a**). This type of behaviour used to be the conventional one [1]. The value of $f_1$ was calculated using the reliable values of the flexoelectric coefficients $f_{ij}$ and tabulated elastic modulus $c_{ij}$. and $\eta=0$. The critical radius $R_{cr}$ (that corresponds to $T=0$) is about 6 nm for both cases $f=0$ and $f=f_1$. The further increase of f value leads to the dramatic change of phase diagram form and transition temperature behaviour (see curves 2-4 in the **Figure 2a**). Note, that the parameter $f$ can be several times higher than the calculated value $2.01 \times 10^{-6}$ V/Pa due to the symmetry change at the surface, so the highest value $f_4=22.12\times10^{-6}$ V/Pa is reasonable. The curve 2 has two critical radiuses, $R_{cr}^{\min}$ and $R_{cr}^{\max}$, the minimal is about 1 nm and the maximal is about 5 nm. It is not clear whether the phenomenological theory is applicable for reliable determination of the minimal critical radii. The phase sequence corresponding to the curve 2 is the ferroelectric (FE) phase at $R > R_{cr}^{\max}(T)$, paraelectric (PE) phase at $R_{cr}^{\min}(T) < R < R_{cr}^{\max}(T)$, and again FE phase at $R < R_{cr}^{\min}(T)$. Here we are faced with the new type of tetragonal ferroelectric phase reentrance with $R$ decrease. Classical reentrant phase without PE phase and $R_{cr}$ occurs when $T_{FE}$ firstly decreases with $R$ decrease and then increase at $R<R_r$ (curves 3 and 4 in the **Figure 2a**). $R_r$ is an analog of a "turning radius" and $T_r$ is the "turning temperature", at that $T_{FE}(R_r) = T_r$. The temperature $T_r$ and radius $R_r$ strongly increases with $f$ increase (compare the curves 3 and 4 in the **Figure 2a**). The transition temperature is insensitive to the sign of the parameter $f$, because its contribution to the constant $B$ is quadratic.

Different curves 1-4 in the **Figure 2b** correspond to zero flexo-parameter ($f=0$) and compressive Vegard strains ($\eta<0$) with the absolute value $\eta_1<\eta_2<\eta_3<\eta_4$. The general form of the phase diagram in **Figs.2a** and **2b** is similar. Actually, the ferroelectric transition temperature monotonically increases with particle radius increase for $\eta=0$ and $f=0$ (curve 1 in the **Figure 2b**). However, quantitatively, the increase of $T_{FE}$ with the particle radius decrease appears under the moderate increase of the compressive strain $\eta$ from 0.4% to 1.5% (see curves 2-4 in the **Figure 2b**). The curve 2 has two critical radiuses, $R_{cr}^{\min} \approx 1.5$ nm and $R_{cr}^{\max} \approx 5.2$ nm. The "turning point" temperature $T_r$ strongly increases with $\eta$ increase (compare the curves 3 and 4 in the plots). Tensile strains $\eta>0$ favour the temperature decrease with the particle radius decrease.

Since both chemical pressure and flexoelectric contributions are additive to the constant $B$ given by Eq.(7b), the curves in the **Figures 2a** and **2b** demonstrate the same trends with either $f$ increase or $\eta$ increase.

Room temperature polarization vs. the particle radius was calculated at different effective flexo-constant $f$ and Vegard strain $\eta$. It is shown in the **Figures 2c-d.** Different curves 1-4 in the **Figure 2c** correspond to zero Vegard strain ($\eta=0$) and different effective flexo-parameter $f_1<f_2<f_3<f_4$ exactly the same as in the **Figure 2a**. Different curves 1-4 in the **Figure 2d** correspond to zero flexo-parameter ($f=0$) and different compressive Vegard strains $\eta_1<\eta_2<\eta_3<\eta_4$ exactly the same as in the **Figure 2b**. Reentrant FE phase is clearly seen for the curves 4. Curves 2 and 3 have two regions of FE phase, separated by PE phase in the radii range $R_{cr}^{min}(T) < R < R_{cr}^{max}(T)$ at $T=293$ K. The critical radii can be defined graphically as the intersection points of the horizontal line $T=293$ K with the curves 2 or 3 in the **Figures 2a-b.** Note that this statement is value for curves 2 at all temperatures and for the curves 3 for $T > T_r$ only. At $T < T_r$ one can expect the polarization behavior similar to the curves 4. Curves 1 has only one PE region at $R < R_{cr}(T)$ and FE region at $R > R_{cr}(T)$ (corresponding curves 1 in the **Figures 2a-b** have only one intersection point wit the horizontal line $T=293$ K.

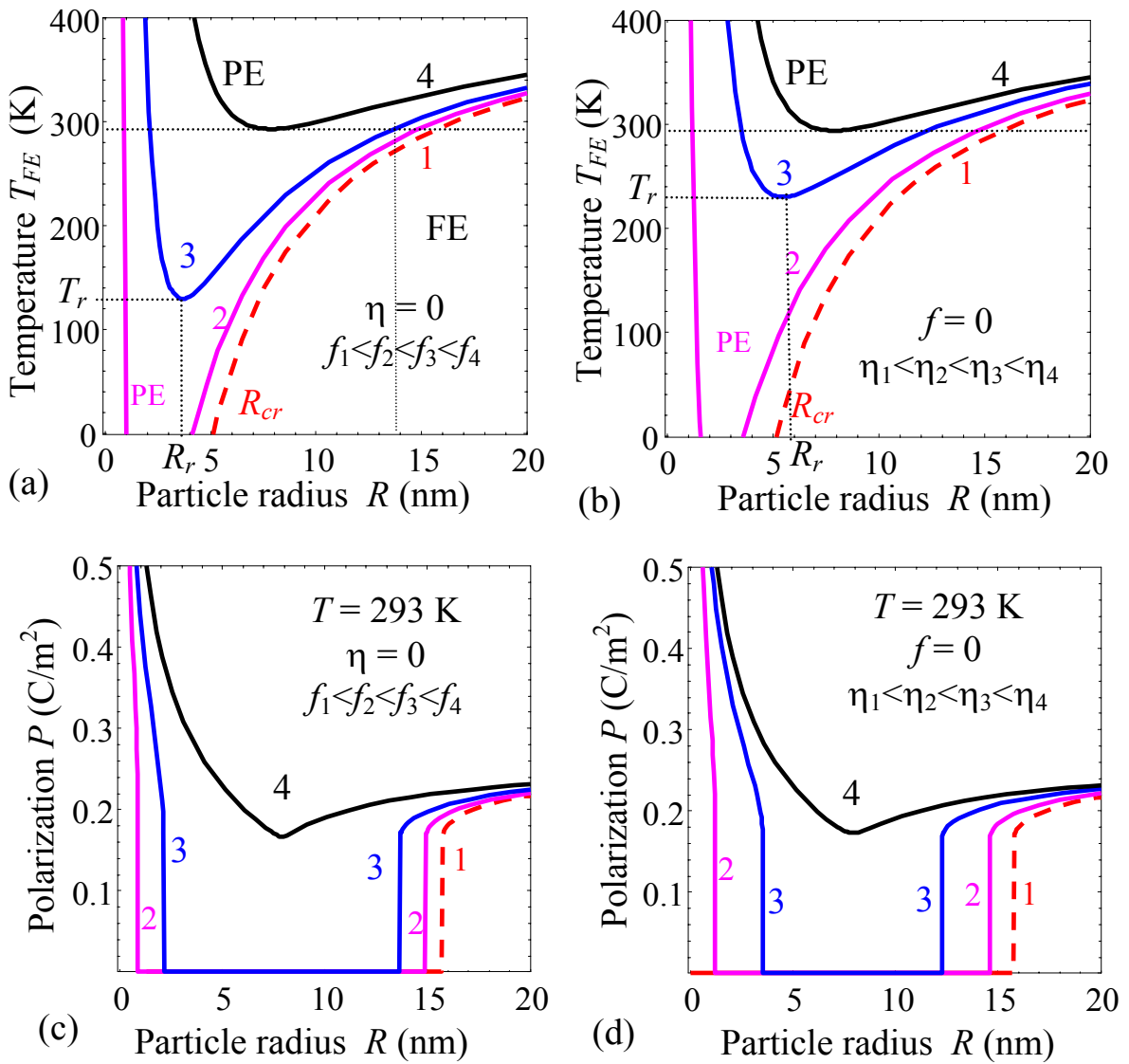

**Figure 2.** (a) Transition temperature $T_{FE}$ vs. the particle radius calculated for zero Vegard strain $\eta=0$ and different effective flexo-parameter $f$: $f_1=2.01\times10^{-6}$ V/Pa (dashed curve 1), $f_2=5f_1$ (solid

curve 2), $f_3=7.5f_1$ (dotted curve 3), $f_4=11f_1$ (solid curve 4). **(b)** Transition temperature $T_{FE}$ vs. the spherical particle radius calculated for zero effective flexo-constant $f=0$ and different Vegard strain η: $η_1=0$ (dashed curve 1), $η_2=-0.004$ (solid curve 2) $η_3=-0.01$ (dotted curve 3), $η_4=-0.0145$ (solid curve 4). **(c)-(d)** Spontaneous polarization vs. the spherical particle radius calculated for different flexo-parameter and Vegard strain values exactly the same as in the plots **(a)** and **(b)**. Temperature T=293 K. Other parameters of BaTiO$_3$ are listed in the **Table SI.**

Tetragonality vs. the particle radius is shown in the **Figures 3a-b.** Different curves 1-4 in the **Figure 3a** correspond to zero Vegard strain (η=0) and different flexo-parameter $f_1<f_2<f_3<f_4$. Different curves 1-4 in the **Figure 3b** correspond to zero flexo-parameter ($f=0$) and different compressive Vegard strains $η_1<η_2<η_3<η_4$.

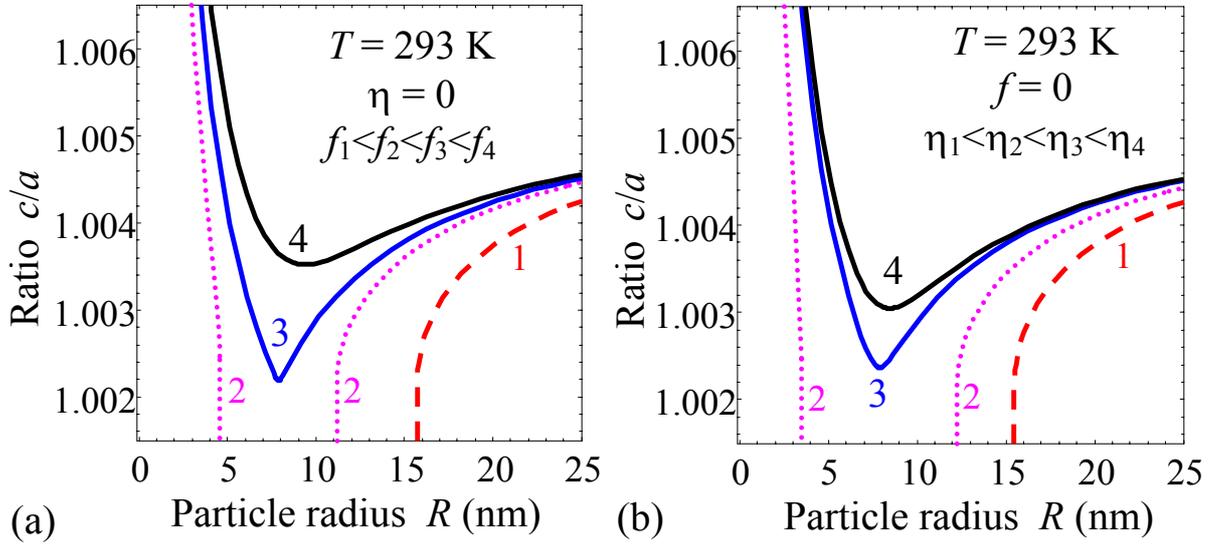

**Figure 3. (a)** Tetragonality vs. the particle radius calculated for zero Vegard strain η=0 and different effective flexo-parameter $f$: $f_1=2.01×10^{-6}$ V/Pa (dashed curve 1), $f_2=10f_1$ (dotted curve 2), $f_3=11f_1$ (solid curve 3), $f_4=12f_1$ (solid curve 4). **(b)** Tetragonality vs. the spherical particle radius calculated for zero flexo-parameter $f=0$ and different Vegard strain η: $η_1=0.001$ (dashed curve 1), $η_2=-0.010$ (dotted curve 2) $η_3=-0.015$ (solid curve 3), $η_4=-0.016$ (solid curve 4). Temperature T=293 K. Other parameters of BaTiO$_3$ are listed in the **Table SI.**

The tetragonality ratio $c/a$ monotonically increases with particle radius increase for the curves 1 corresponding to zero or smallest $f$ and η. The increase of $c/a$ ratio with the particle radius decrease appears under the increase of either flexo-parameter $f$ or compressive strains η (see curves 2-4). The curve 2 has two critical radiuses, $R_{cr}^{min} \approx (3.5-5)$ nm and $R_{cr}^{max} \approx (11-$

12) nm. The "turning point" temperature $T_r$ strongly increases with $f$ or $\eta$ increase (compare the curves 3 and 4). Reentrant FE phase is clearly seen for the curves 3 and 4.

**Figure 4a** shows the best fitting of our theory (solid curve) to experimental data [21] (symbols with error bars) achieved for the definite values of the parameters $A = 1.6079$ K× $m^2J/C^2$ and $B = -6.81869 \times 10^{-9}$ K×$m^3J/C^2$ in Eq.(6). We did not aim to fit well the tetragonality for the smallest particle of radius 2.5 nm, because here the phenomenological continuous approach may be invalid. The fitting procedure does not allow us to extract the values $\eta$ and $f$ separately, but only their combination in the constant $B$.

**Figures 4b** and **4c** illustrate the dependences of the spontaneous polarization and transition temperature on the spherical particle radius calculated for the same parameters $A$ and $B$ than in the **Figure 4a**. So that we "reconstruct" the polarization and transition temperature from to the best fitting of the tetragonality measured experimentally [21]. The reconstructed dependences clearly demonstrate the strong (more than 2 times) enhancement of polarization and transition temperature for the particle radius less than 10 nm. The reentrant phase region appears for radii $R<10$ nm. One can see that theoretical Figure 2 describe pretty good the spontaneous polarization and transition temperature reconstructed from experimental $c/a$ value.

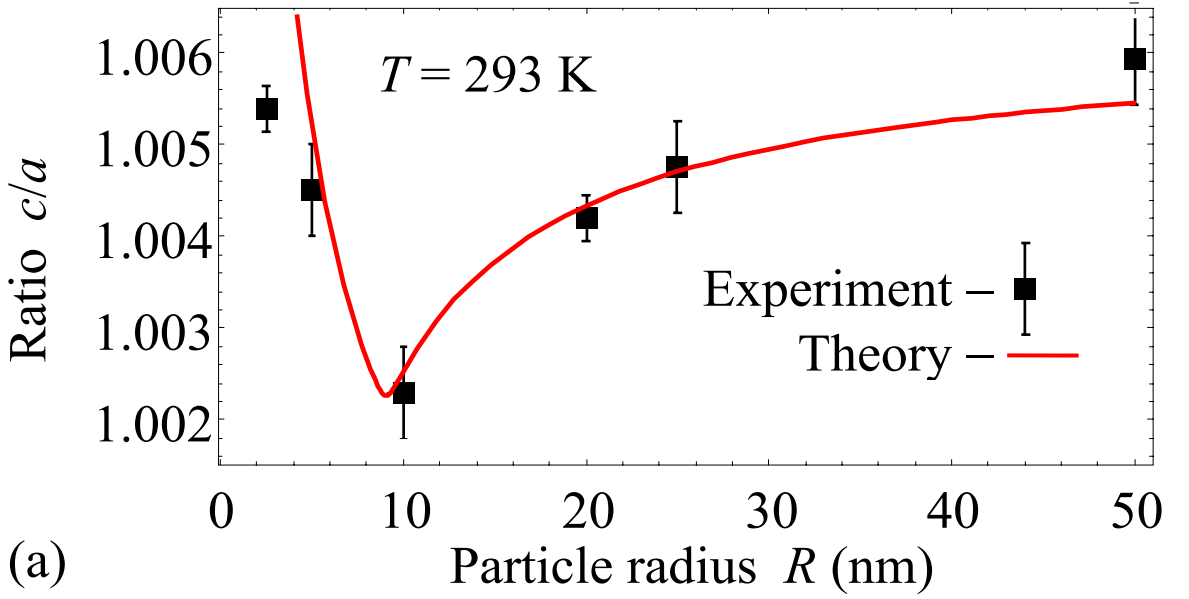

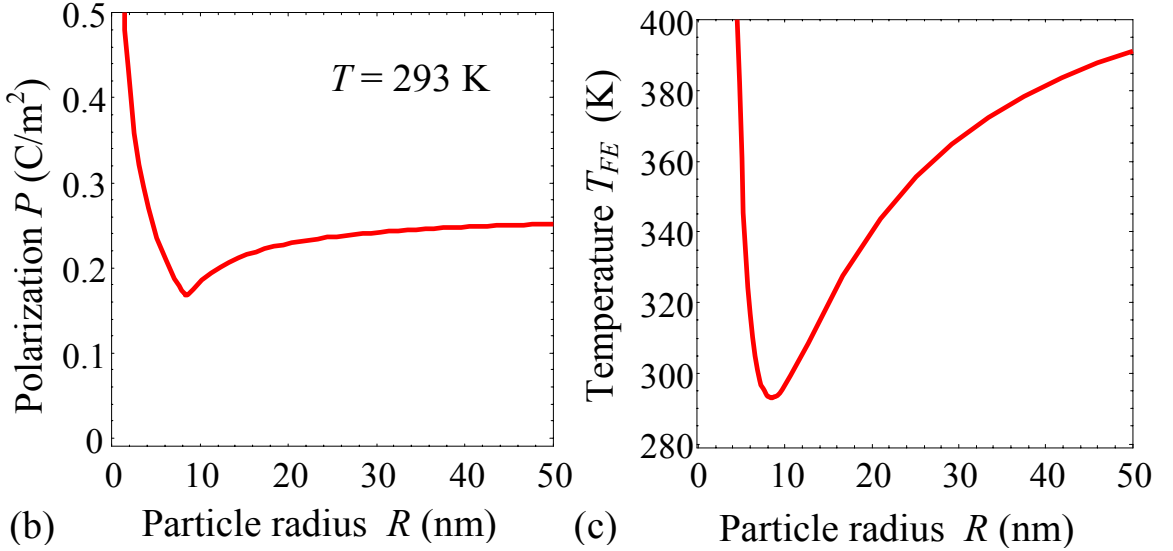

**Figure 4.** Room temperature tetragonality ratio c/a **(a)**, spontaneous polarization **(b),** and transition temperature **(c)** vs. the particle radius**.** Plot **(a)** shows the best fitting of our theory (solid curve) to experimental data [21] (symbols with error bars) achieved for the values of the constants $A$ = 1.6079 K×m$^2$J/C$^2$ and $B$ = −6.81869×10$^{−9}$ K×m$^3$J/C$^2$ in Eq.(6). T=293 K. Other parameters of BaTiO$_3$ are listed in the **Table SI.**

### IV. Discussion

It follows from the **Figures 2-4** that both considered mechanisms, namely the spontaneous flexoeffect and Vegard mechanism reflecting the influence of elastic dipoles originated from defects (in the oxides the defects are mainly the oxygen vacancies), described experimental points good enough, although the best fitting was obtained by the mixture of the mechanisms (**Fig.4**). However up to now the value of the coefficients $f$ and $\eta$ for nanostructured BaTiO$_3$ were neither measured nor calculated on the basis of the first principles approach.

Because of this it seems reasonable to propose a method that can decrease one of these two mechanisms contribution. While the spontaneous flexoelectric effect exists in any nanostructured ferroelectrics, the contribution of oxygen vacancies can be decreased by annealing in oxygen atmosphere similarly to the procedure used to find out mechanism of $3d^0$ magnetization in nano-oxides (see e.g. [58, 59, 60, 61, 62]). It can be expected that the annealing should decrease essentially Vegard mechanism contribution and to obtain the spontaneous flexoeffect coefficient experimentally for the first time. It is worth to underline that it could be essential size dependence of flexocoefficient, because general properties of flexoelectricity and piezoelectricity are similar and their estimated value is *e/a*, where *a* is characteristic atomic separation [24, 36]. Strong size dependence of piezoelectric coefficient made it possible to explain large value of magnetoelectric effect in some nanostructured multiferroics [63]. Moreover it was shown [1] that piezoelectric tensor components related to unit cell deformation is proportional to flexoeffect coefficient. Possible size dependence of the flexoelectric coefficient will open the way to increase the effective flexoelectric coefficient and so its contribution to the appearance of reentrant phases in any nanoferroelectric.

On the other hand investigation of the phase diagrams in other nanostructured perovskites similar to $BaTiO_3$, e.g. ferroelectric $KTa_{1-x}Nb_xO_3$ (x>0.05) with different coefficients f and η in comparison in $BaTiO_3$, and so with different number of $R_{cr}$ and therefore the type of the phase diagrams, namely FE-PE at some $R_{cr}$, reentrant FE phase without PE and any $R_{cr}$, FE to PE phase transition at $R_{cr}^{max}$ and PE to FE transition at $R_{cr}^{min}$. The theoretical forecast is waiting for experimental verification.

## V. Conclusion

Using Landau-Ginzburg-Devonshire phenomenological approach we established the impact of the flexoelectric and Vegard effect on the phase diagrams, long-range polar order and related physical properties of the spherical ferroelectric nanoparticles. The synergy of these effects can lead to the remarkable changes of the nanoparticles' phase diagrams. In particular, a commonly expected transition from ferroelectric to paraelectric phase at some small critical size is absent; so that the critical size loses its sense. The stabilization of the ferroelectric phase manifests itself by the enhancement of the transition temperature and polarization with the particle size decrease (reentrant ferroelectric phase). Appeared that both flexoelectric and Vegard effects, abbreviated as ***flexo-chemo***, can be a decisive physical mechanism ruling the observed phenomena. Since the spontaneous flexoelectric effect, as well as ion vacancies, should exist in any nanostructured ferroelectrics, obtained analytical results can be valid for many nanoferroelectrics, where reentrant phases appearance can be forecasted.

**Acknowledgements**

A.N.M. acknowledges National Academy of Sciences of Ukraine (grant 35-02-15 and joint Ukraine-Belarus grant 07-06-15). Authors acknowledge useful remarks and multiple discussions with E.A. Eliseev (NASU).

**Authors' contribution**

M.D.G. generated the idea that the spontaneous flexoelectric effect can induce the reentrant phase in nanoferroics; she is the main contributor to the introductive and discussion part of the manuscript. A.N.M. performed the analytical calculations of the flexo-chemical mechanism impact, generated figures and compared theory with experiment; she wrote the original part of the manuscript.

# Supplementary Materials to "Reentrant phase in nanoferroics induced by the flexoelectric and Vegard effects"

Anna N. Morozovska[1] and Maya D. Glinchuk [2*],

[1] Institute of Physics, National Academy of Sciences of Ukraine,
46, pr. Nauky, 03028 Kyiv, Ukraine

[2] Institute of Problems for Material Sciences, National Academy of Sciences of Ukraine,
3, Krjijanovskogo, 03068 Kyiv, Ukraine

**Appendix A**

The equations of state $\delta\Phi_b/\delta P_3 = 0$ and $\delta\Phi_b/\delta\sigma_{ij} = -u_{ij}$ ($\delta$ is the symbol of variation derivative) are obtained by variation of the bulk energy density functional (1) in the particle core:

$$\left(\alpha_b(T) - 2Q_{ij33}\sigma_{ij}\right)P_3 + \beta P_3^3 + \gamma P_3^5 + g_{ij33}\left(\frac{\partial^2 P_3}{\partial x_i \partial x_j}\right) = F_{3ijk}\frac{\partial\sigma_{ij}}{\partial x_k} + E_3^d + E, \qquad (A.1a)$$

$$u_{ij} = s_{ijkl}\sigma_{kl} + Q_{ij33}P_3^2 - F_{ij3l}\frac{\partial P_3}{\partial x_l} + W_{ij}\delta N. \qquad (A.1b)$$

Here $\sigma_{jk}$ is the stress tensor, $u_{jk}$ is the strain tensor. The equations should be solved along with the equations of mechanical equilibrium $\partial\sigma_{ij}(\mathbf{x})/\partial x_i = 0$ and compatibility relation equivalent to the mechanical displacement vector $u_i$ continuity [1]. Variation of the surface and bulk free energy (1) on $P_i$ yields to Euler-Lagrange equations with the boundary conditions [2]:

$$\left(g_{kjim}n_k\frac{\partial P_m}{\partial x_j} + a_{ij}^S P_j + \frac{F_{jkim}}{2}\sigma_{jk}n_m\right)\bigg|_S = 0. \qquad (A.2)$$

Here $\vec{n} = (\vec{e}_r, 0, 0)$ is the external normal to the particle spherical surface, polarization in spherical coordinates $P_r = P_S\cos\theta$, $P_\theta = -P_S\sin\theta$, $P_\varphi = 0$. The most evident consequence of the flexo-coupling is the inhomogeneous boundary conditions. We consider *mechanically free* nanoparticles without misfit dislocations.

The flexoeffect leads to the renormalization of the gradient coefficients $g_{11}^* = g_{11} - f_{11}^2/c_{11}$, $g_{12}^* = g_{12} - f_{44}^2/c_{44}$ and extrapolation length $\lambda^* = \lambda\left(1 - \frac{f_{11}^2}{2c_{11}g_{11}}\right)$, where $\lambda = g_{11}/\alpha^S$.

---

[*] Corresponding author: glim1@voliacable.com

Hereinafter $c_{ij}$ are elastic stiffness, $f_{ij}$ are the flexoelectric strain coefficients. The renormalized gradient coefficients should be positive for the system stability, and the stability conditions $g_{11}^* > 0$ and $g_{12}^* > 0$ determine the upper limits for the values of flexoelectric coefficients (compare with Ref.[3]). The characteristic length, $R_z = \sqrt{g_{11}^* \varepsilon_0 \varepsilon_b}$.

The parameter $f \cong \dfrac{g(f_{12}c_{11} - c_{12}f_{11})}{\sqrt{c_{11}(c_{11}^2 + c_{11}c_{12} - 2c_{12}^2)}}$ is proportional to the flexocoupling constants $f_{ij}$, elastic stiffness $c_{ij}$ and geometrical dimensionless factor $g$. For estimations the expression for $f$ for a sphere is regarded two times different from the one derived for the pills in Ref.[2]. It is worth to underline that $f$ values can be different from the bulk ones due to the symmetry change at the surface [4].

### Appendix B

The monotonic decrease of the transition temperature with $R$ decrease appears under the condition $A>0$ and $B>0$, because the size-dependent contribution $\left(\dfrac{A}{R} + \dfrac{B}{R^2}\right)$ is positive and monotonically increases with $R$ decrease, so that $T_{FE}(R)$ decreases. Corresponding critical radius defined from the condition $T_{FE}(R_{cr}) = 0$ is equal to $R_{cr} = \dfrac{\sqrt{A^2 + 4BT_\theta^b \alpha_T} + A}{2T_\theta^b \alpha_T}$ (here $A>0$ and $B>0$).

The increase of the transition temperature for some $R$-range can appear in the case $B<0$ under the condition $\left(\dfrac{A}{R} + \dfrac{B}{R^2}\right) < 0$. Since the first term $A/R$ is always positive, it favors the nanosphere transition into the non-polar paraelectric cubic phase, correspondingly decreases the spontaneous polarization and tetragonality $c/a$ with $R$-decrease. When the second term $B/R^2$ is negative and thus increases it the transition temperature, enhances polar properties and tetragonality with $R$ decrease, since $1/R^2$ is stronger that $1/R$ at small $R$. The competition of these two terms results into the appearance of two critical radiuses, $R_{cr}^{\min} = \dfrac{A - \sqrt{A^2 + 4BT_\theta^b \alpha_T}}{2T_\theta^b \alpha_T}$ and $R_{cr}^{\max} = \dfrac{\sqrt{A^2 + 4BT_\theta^b \alpha_T} + A}{2T_\theta^b \alpha_T}$, which exist under the conditions $A>0$, $B<0$ and $A^2 + 4BT_\theta^b \alpha_T > 0$. At that $R_{cr}^{\min}$ tends to zero at $A >> 2\sqrt{-BT_\theta^b \alpha_T}$. Under the condition $A^2 + 4BT_\theta^b \alpha_T < 0$ the critical radiuses do not exist and the nanoparticle maintains its ferroelectric state up to the ultra-small sizes. Exactly the reentrant ferroelectric phase can appear in the case and enhances at small

*R*. Since both conditions $B<0$ and $A^2 + 4BT_0^b\alpha_T < 0$ can be valid for relatively high flexo-parameter $f^2$ and/or compressive Vegard strains η<0, only these two mechanisms can enhance the ferroelectric properties of nanospheres (in contrast to the nanowires, where the surface tension can maintain and improve the properties due to the condition $Q_{12} < 0$ [**Ошибка! Закладка не определена.**]).

## Appendix C

**Table SI.** Material parameters used in the simulations

| coefficient | ferroelectric BaTiO$_3$ (collected and recalculated mainly from Ref. [a, b]) |
|---|---|
| **Symmetry at room *T*** | tetragonal |
| ε$_b$ (the same is shell) | 7 (Ref. [b]) |
| α (C$^{-2}$·mJ) | 6.68(*T*–381)×10$^5$ |
| β$_{ij}$ (C$^{-4}$·m$^5$J) | β$_{11}$= 18.76(*T*–393)×10$^6$–8.08×10$^8$, β$_{12}$= 12.92×10$^8$ |
| γ$_{ijk}$ (C$^{-6}$·m$^9$J) | γ$_{111}$= –33.12(*T*–393)×10$^7$+16.56×10$^9$, γ$_{112}$=26.82×10$^9$, γ$_{123}$=29.46×10$^9$ |
| Q$_{ij}$ (C$^{-2}$·m$^4$) | Q$_{11}$=0.11, Q$_{12}$= –0.043, Q$_{44}$=0.059 |
| s$_{ij}$ (×10$^{-12}$ Pa$^{-1}$) | s$_{11}$=8.3, s$_{12}$= –2.7, s$_{44}$=9.24 |
| c$_{ij}$ (×10$^{11}$ Pa) | c$_{11}$=1.76, c$_{12}$= 8.46, c$_{44}$=1.08 |
| g$_{ij}$ (×10$^{-10}$C$^{-2}$m$^3$J) | g$_{11}$=5.1, g$_{12}$= –0.2, g$_{44}$= 0.2 [c], g$_{11}^*$=3.6 |
| f$_{ijkl}$ (V) | ~100 (estimated from measurements of Ref. [d])<br>f$_{11}$ = 5.12, f$_{12}$ =3.32, f$_{44}$ ≈ 0.045 ± 0.015 [e] |
| F$_{iklj}$ (×10$^{-11}$C$^{-1}$m$^3$) | ~100 (estimated from measurements of Ref. [d])<br>F$_{11}$= +2.46, F$_{12}$=0.48, F$_{44}$=0.05 (recalculated from [e] using F$_{αγ}$=f$_{αβ}$s$_{βγ}$) |
| R$_z$ (nm) | R$_z$ = 0.15 |
| R$_0$ (nm) | R$_0$= 0.8 - 4 (reasonable interval is 0.4 – 5) |
| R$_S$ (nm) | R$_S$ = 0.1 (reasonable interval is 0.05 – 2) |
| f (×10$^{-6}$ V/Pa) | 2.01 – 20.01 (reasonable interval for the bulk material is 1 – 8) |
| λ (nm) | <0.1 (reasonable interval is 0 – 4) |
| μ(N/m) | 1.5 (reasonable interval is 1 – 3) |
| η | 0.001 - 0.015 (reasonable interval of dimensionless strain 0.001 – 0.03) |

[a] A.J. Bell. J. Appl. Phys. 89, 3907 (2001).

[b] J. Hlinka and P. Márton, Phys. Rev. B 74, 104104 (2006).

[c] P. Marton, I. Rychetsky, and J. Hlinka. Phys. Rev. B 81, 144125 (2010).

[d] W. Ma and L. E. Cross, Appl. Phys. Lett., 88, 232902 (2006).

[e] I. Ponomareva, A. K. Tagantsev, L. Bellaiche. Phys.Rev B 85, 104101 (2012).